\documentclass[sigconf]{acmart}

\usepackage{booktabs} 

\DeclareMathOperator*{\argmax}{arg\,max}

\usepackage{algorithm}
\usepackage{algorithmic}

\usepackage{multirow}

\usepackage{subcaption}

\setcopyright{none}

\acmDOI{}

\acmISBN{}

\acmConference[KDD'17]{ACM SIG KDD}{August 2017}{Halifax, Nova Scotia Canada} 
\acmYear{2017}
\copyrightyear{2017}

\newcommand{\name}{GraphZip}
\newcommand{\system}{\textsc}
\newcommand{\batchsize}{$\alpha$}
\newcommand{\dictsize}{$\theta$}

\begin{document}
\title{\system{\name}: Dictionary-based Compression 
for Mining Graph Streams}

\author{Charles A. Packer}
\affiliation{%
  \institution{University of California, San Diego}
  \city{La Jolla} 
  \state{California} 
}
\email{cpacker@cs.ucsd.edu}

\author{Lawrence B. Holder}
\affiliation{%
  \institution{Washington State University}
  \city{Pullman} 
  \state{Washington} 
}
\email{holder@wsu.edu}








\begin{abstract}
A massive amount of data generated today on platforms such as social networks, telecommunication networks, and the internet in general can be represented as graph streams. 
Activity in a network's underlying graph generates a sequence of edges in the form of a stream; for example, a social network may generate a graph stream based on the interactions (edges) between different users (nodes) over time. 
While many graph mining algorithms have already been developed for analyzing relatively small graphs, graphs that begin to approach the size of real-world networks stress the limitations of such methods due to their dynamic nature and the substantial number of nodes and connections involved.\\ 
\indent In this paper we present \system{\name}, a scalable method for mining interesting patterns in graph streams. \system{\name} is inspired by the Lempel-Ziv (LZ) class of compression algorithms, and uses a novel dictionary-based compression approach in conjunction with the minimum description length principle to discover maximally-compressing patterns in a graph stream. We experimentally show that \system{\name} is able to retrieve complex and insightful patterns from large real-world graphs and artificially-generated graphs with ground truth patterns. Additionally, our results demonstrate that \system{\name} is both highly efficient and highly effective compared to existing state-of-the-art methods for mining graph streams.
\end{abstract}

%
%




\maketitle

\begin{figure}
\includegraphics[width=3.2in]{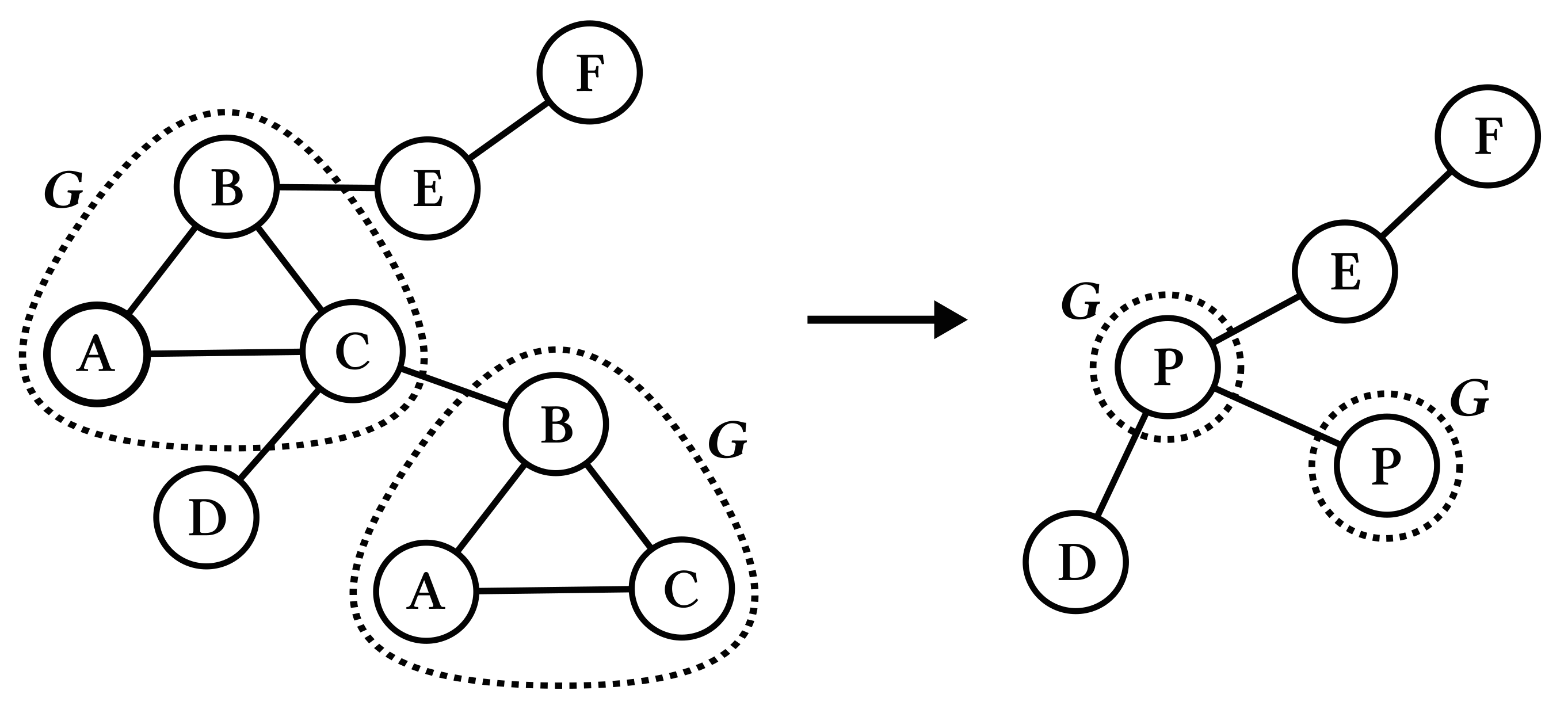}
    \caption{Compression via substitution. Vertices $A$, $B$, and $C$ form a recurring pattern (subgraph) $G$. Substituting the pattern for a single node $P$ representing $G$ reduces the graph's overall size. The reduction in size is a measure of the compression afforded by pattern $P$. 
    }\label{fig:compress}
\end{figure}

\begingroup
\section{Introduction}
Graphs are used to represent data across a wide spectrum of areas, from computational chemistry 
to social network analysis.
Graph mining is an active 
area of research, and there are numerous methods for mining smaller graphs (several thousand edges), but many of these systems are unable to scale to real-world graphs (e.g., social networks) with millions or even billions of edges. 
Conventional graph mining algorithms assume a complete static graph as input, however many real-world graphs are often too large to hold in main memory. 
Additionally, many real-world graphs of interest are dynamic and actively growing - Facebook, for example, records over 300 new users per minute and has a social graph with more than 400 billion edges  \cite{ChingFB}. 
While it is possible to utilize conventional graph mining systems on dynamic graphs by processing static `snapshots' of the graph at various points in time, 
in many cases the underlying data the graph represents changes at a rate so fast that attempting to analyze the data using such methods is futile.

In cases where the graph in question is inherently dynamic, we can instead treat the graph as a sequential stream of edges representing continuous updates to the graph's overall structure. 
For example, given a graph modeling friendships (edges) between users (nodes) in a social network, we can consider all new or updated relationships during a set time interval (e.g., 1 hour) a set of edges from time $t_i$ to $t_{i+1}$. The graph mining system then processes the sequential edge sets at every interval, as opposed to attempting to read the entire graph at once. Processing large graphs in a streaming fashion drastically reduces the system's memory requirements (since only small portions of the graph are seen at a time) and enables processing of large, dynamic real-world datasets. 
However, deploying a streaming model for real-time data analysis also imposes strict constraints: the system has a limited time window to process each set of edges, and edges can only be viewed once before they are replaced in memory by those in the next set. 

Many graph mining algorithms aim to identify interesting patterns within an input graph. Various algorithms use different metrics to quantify how `interesting' a pattern is: frequent subgraph mining (FSM) focuses on finding all subgraphs that appear in the graph over a certain frequency threshold, whereas problems such as counting motifs or finding maximal cliques in a graph (formalized in \cite{przulj-motifs} and \cite{bron-kerbosch-cliques}, respectively) focus on discovering subgraphs with a specific structure. 

This paper relies on a novel approach to identify interesting patterns in a graph - namely, finding a set of substructures that best compress the graph.
More precisely, we compress a graph using a pattern (subgraph) $G$ by replacing all instances of $G$ in the graph with a new node $p$ representing $G$
(see figure \ref{fig:compress}). The reduction in size of the overall graph is a measure of the compression afforded by pattern $G$, and we search for patterns that compress the graph to the maximal extent. 
The same concept is found in certain types of data compression (e.g., LZ78 \cite{lz78}, ZIP) where the compression method looks for recurring patterns or sequences in the data stream, builds a dictionary representing the recurring patterns with shorter binary codes, and then stores the compressed data using only the binary codes and the dictionary. In the context of graphs, a byproduct of this process is that the pattern dictionary contains a set of subgraphs that compress well, and therefore represents an alternative approach for finding interesting (highly-compressing) patterns in a graph stream.

We propose a dictionary-based compression method for graph-
based knowledge discovery: \system{\name}. Our approach is designed to efficiently mine graph streams and uncover interesting patterns by finding maximally-compressing substructures. Specifically, our main contributions are as follows:

\begin{enumerate}
\item We propose a new graph mining paradigm based on the LZ class of compression algorithms.
\item Based on this paradigm, we introduce a new graph mining algorithm, \system{\name}, for efficiently processing massive amounts of data from graph streams.
\item We demonstrate the effectiveness and scalability of our method using a variety of openly available synthetic and real-world datasets.
\end{enumerate}

In our experiments, 
we demonstrate that our approach is able to retrieve both complex and insightful patterns from large real-world graphs by utilizing graph streams. 
In addition, 
we show that 
our approach is able to successfully mine a large class of varied substructures from artificially-generated graphs with ground truth patterns. 
When we compare 
\system{\name}'s performance with that of several other state-of-the-art graph mining methods, 
we find that 
\system{\name} consistently outperforms state-of-the-art methods on a variety of real-world datasets. 

The \system{\name} system, including all related code and data used for this paper, are available for download online\footnote{\url{https://github.com/cpacker/graphzip}}.
\system{\name} is not to be confused with the method described in \cite{graphzip2} for hierarchical clustering on spatial data, which goes by the same name.
\endgroup

\section{Related Work}
For the purposes of this paper, we classify previous work into two general categories: streaming and non-streaming.




\subsection{Non-streaming}
Non-streaming graph mining algorithms take as input either a single graph (single graph mining), or multiple smaller graphs (transactional mining).

\textbf{Transactional mining.}
\system{FSG} \cite{fsg} is an early approach to finding frequent subgraphs across a set of graphs, and adopts the Apriori algorithm for frequent itemset mining \cite{apriori}. \system{FSG} works by joining two frequent subgraphs to construct candidate subgraphs, then checking the frequency of the new candidates in the graph.
\system{gSpan} \cite{gspan} uses a `grow-and-store' approach 
that extends saved subgraphs to form new ones, an improvement over \system{FSG}'s prohibitively expensive join operation.
\system{Margin} \cite{margin} prunes the search space to find maximal subgraphs only, a narrower and thus easier problem than FSM.
\system{CloseGraph} \cite{closegraph} is another method that reduces the problem space by mining only \emph{closed} frequent subgraphs - subgraphs that have strictly smaller support than any existing supergraphs.
\system{Leap} \cite{leap} 
and \system{GraphSig} \cite{graphsig} are two recent approaches for mining `significant subgraphs' as measured by a probabilistic objective function. By mining a small set of statistically significant subgraphs as opposed to a complete set of frequent subgraphs, \system{Leap} and \system{GraphSig} are able to avoid the problem of exponential search spaces generated by FSM miners with low frequency thresholds.

\textbf{Single graph mining.}
\system{SUBDUE} \cite{subdue} is an approximate algorithm based on the branch-and-bound search technique. \system{SUBDUE}, like \system{\name}, uses the Minimum Description Length (MDL) principle \cite{mdl} to mine maximally-compressing patterns in the graph. However, unlike \system{GraphZip}, \system{SUBDUE} 
returns a restrictively small number of patterns regardless of the size of the input graph \cite{grew}. \system{SUBDUE} has been improved in recent years \cite{subdue2-subgen,subdue3,subdue4}, yet the fundamental limitations of the algorithm (in particular the branch-and-bound technique) remain the same.
\system{SEuS} \cite{seus} is an approximate method 
that creates a compressed representation of the graph by collapsing vertices that share labels. \system{SEuS} however is only effective in cases where the input graph has a small number of unique subgraphs that occur with high frequency, as opposed to when the input graph has a large number of subgraphs that appear with lower frequency. 
\system{SiGram} \cite{sigram} is a complete method for finding frequent connected subgraphs 
(complete methods are guaranteed to find all solutions that fit certain constraints such as the minimum frequency threshold, unlike their approximate counterparts). \system{SiGram} adopts a grow-and-store approach similar to \system{gSpan}, but uses the expensive (NP-complete) Maximal Independent Set (MIS) metric for its frequency threshold, leading the system to be comparatively inefficient in practice. Additionally, like \system{SEuS}, \system{SiGram} suffers from a limited domain problem as it is designed specifically to mine sparse, undirected and labeled graphs only. 
\system{Grew} \cite{grew} 
is another approximate method for mining frequent connected subgraphs, and is similar to \system{SUBDUE} in that \system{Grew} only discovers a relatively small subset of solutions in the search space. 
\system{GraMi} \cite{grami} is a state-of-the-art complete method (with an approximate version \system{AGraMi}) that has been shown to be highly-efficient for FSM on a single large graph. However, the size of the input graph is still 
limited since \system{GraMi} requires the entire graph to be held in main memory.
\system{Arabesque} \cite{arabesque} is a recent distributed approach built on top of \system{Apache Giraph} \cite{giraph} that can horizontally scale non-streaming algorithms (FSM, clique finding, motif counting, etc.) across multiple servers. However, horizontal scaling can be cost-prohibitive and is only capable of linearly scaling algorithms whose runtimes often grow exponentially with the size of the input graph. 
\system{GERM} \cite{germ} and the algorithm introduced by Wackersreuther et al. \cite{wackersreuther} can mine frequent subgraphs in dynamic graphs, however both methods require as input snapshots of the entire graph as opposed to incremental updates to the graph via graph streams.

\subsection{Streaming}
\system{GraphScope} \cite{graphscope} is a parameter-free streaming method that, like \system{GraphZip} and \system{SUBDUE}, is based on the MDL principle. \system{GraphScope} encodes the graph stream with the objective of minimizing compression cost, in order to determine important change-points in the temporal data. Beyond change-point and community detection however, \system{GraphScope} has limited use for other tasks, e.g. mining interesting subgraphs. Though the model itself is parameter-free, \system{GraphScope} requires the dimensions of the graph (number of source and destination nodes) to be known \emph{a priori}, and thus is unable to mine streams from dynamic graphs that introduce unseen nodes in new edge streams. 
Braun et al. \cite{braun2014} proposed a novel data structure called DSMatrix for mining frequent patterns in dense graph streams, yet similar to \system{GraphScope} their approach requires that the edges and nodes be known beforehand, limiting its real-world applications.
Aggarwal et al. \cite{aggarwalvldb2010} introduced a probabilistic model for mining dense structural patterns in graph streams, however the approximation techniques used lead to the occurrence of both false positives and false negatives in the results set, reducing the method's viability in many real-world settings. 
\system{StreamFSM} \cite{streamfsm}, based on \system{gSpan}, is a recently introduced method for frequent subgraph mining on graph streams, whose performance we compare directly with that of \system{\name} (see section \S 4).
There also exist several systems targeted at more specific graph analysis tasks in the streaming setting: counting triangles \cite{doulion}, outlier \cite{aggarwalicde2011} and hotspot \cite{aggarwalicdm2013} detection, and link prediction \cite{zhaoicde2016}. 
Summarization methods such as \system{TCM} \cite{tcm}, \system{gSketch} \cite{gsketch} and \system{count-min sketch} \cite{countminsketch} focus on constructing sketch synopses from large graph streams that can provide approximate answers to queries about the graph's properties. 
For a detailed survey of state-of-the-art graph stream techniques, see \cite{streamsurvey} (for a more general overview of graph mining algorithms, see \cite{gmsurvey}). 

\system{\name} can process an infinite stream of edges without requiring details about nodes or edges beforehand, 
and has no restrictions on the type of graph being streamed. While \system{\name} is designed specifically for the streaming setting, it draws from ideas such as grow-and-store and the MDL principle originally applied in non-streaming methods. In contrast to summarization methods, \system{\name} returns exact subgraphs extracted from the stream as opposed to approximate results. To the best of our knowledge, \system{\name} is the first graph mining algorithm for 
mining maximally-compressing subgraphs from a graph stream.

\begin{table}
  \caption{Symbol definitions. Note that \batchsize{} and \dictsize{} are the hyperparameters of the \system{\name} algorithm.}
  \label{table:1}
  \begin{tabular}{@{}ll@{}}
    \toprule
    Symbol & Definition\\
    \midrule
    $G$ & Arbitrary graph\\ 
    $V_G$ & Vertex set of graph G\\ 
    $E_G$ & Edge set of graph G\\ 
    $S$ & Graph stream sequence\\ 
    $S^{(i)}$ & Graph at time $i$ of stream $S$\\ 
    $B$ & A batch of edges from graph stream\\
    $P$ & Pattern dictionary\\ 
    $P^{(i)}$ & Pattern (graph) $i$ in dictionary $P$\\
    $V_{P^{(i)}}$ & Vertex (node) set of pattern $P^{(i)}$\\
    \addlinespace[0.2em]
    $E_{P^{(i)}}$ & Edge set of pattern $P^{(i)}$\\
    \addlinespace[0.2em]
    $C_{P^{(i)}}$ & Compression score of pattern $P^{(i)}$\\
    \addlinespace[0.2em]
    $F_{P^{(i)}}$ & Frequency (count) of pattern $P^{(i)}$\\
    \batchsize & Batch size\\ 
    \dictsize  & Size threshold of $P$\\ 
    $H(G)$ & Compression scoring function\\
    $I(G, G)$ & Graph isomorphism function\\
    $SI(G, G)$ & Subgraph isomorphism function\\
  \bottomrule
\end{tabular}
\end{table}

\section{Method}

\subsection{Preliminaries}
In this section, we review the fundamental graph theory needed to formulate our approach and formalize the definitions used in the rest of the paper. See table \ref{table:1} for symbol definitions.

\textbf{Terminology.} A graph $G$ is composed of a vertex set $V$ which contains all vertices (nodes) $v \in V$, and an edge set $E$ which contains all edges $e \in E$, each of which connects a source vertex to a target vertex. A subgraph $g$ of $G$ is a graph composed of a subset of $G$'s vertices and edges.
All vertices $v \in V$ and edges $e \in E$ have a unique \emph{index} which refers to its internal location in the edge or vertex list (e.g., $v_1$ in $V = \{v_1, v_2, v_3\}$ has index $0$, $v_2$ has index 1, etc.). In a vertex-labeled graph, there exists a one-to-one (i.e., unique) mapping from each vertex to a \emph{label}, and in an edge-labeled graph the same mapping exists for the edge set. The value of labels within a graph is often domain-dependent: e.g., in a social network, vertex labels may correspond to a user type (e.g. `male', `female') while edge labels may correspond to different relationship types (e.g. `friend', `family', etc.).


\begin{definition} 
\emph{Isomorphism}: 
Two graphs $G_1$ and $G_2$ are isomorphic (denoted by $G_1 \simeq G_2$) if there is a one-to-one mapping between the edges and vertices of $G_1$ and $G_2$. That is, each vertex $v$ in $G_1$ is mapped to a unique vertex $u$ in $G_2$, the two of which must share the same edges, i.e., be adjacent to the same vertices (if the graph is labeled, the vertices and edges must also share the same labels). 
$G_1 \simeq G_2$ is equivalent to $G_1$ and $G_2$ sharing the same structure.
\end{definition}

\begin{definition}
\emph{Subgraph isomorphism}:
Graph $G_1$ is considered a subgraph isomorphism of graph $G_2$ if it is an isomorphism of some subgraph $g_2$ of $G_2$. The actual instance of $g_2$ is called an \emph{embedding} of $G_1$ in $G_2$. The subgraph isomorphism problem is a generalization of the graph isomorphism problem, and is known to be NP-complete \cite{subiso-npcomplete} (unlike the graph isomorphism problem, the complexity of which is undetermined). Despite the problem's complexity, many graph mining algorithms make heavy use of subgraph isomorphism checks for graph matching, and accordingly several optimizations have been made in the past decade which have significantly improved the efficiency of isomorphism (or subgraph isomorphism) checks in practice. 
\end{definition}

\begin{definition}
\emph{Graph stream:}
A graph stream $S$ can be represented as a chronological sequence of edges drawn from a graph. 
\begin{displaymath}
S = \{ e_{(1)}, e_{(2)}, e_{(3)}, ... , e_{(n)} \}
\end{displaymath}
We can process the graph stream by segmenting the stream into distinct sets 
of edges, each set forming a single (possibly disconnected) graph stream object. 
In the case of a dynamic graph,  
updates to the graph can be viewed as new stream objects. 
In the rest of the paper we also refer to graph stream objects as \emph{batches}, where \emph{batch size} refers to the size of the stream object's edge set (i.e., the number of edges in the batch).
\end{definition}

\subsection{Problem Formulation}

Given a graph stream and a compression scoring function $H$, our objective function equates to maximizing the cumulative compression score of the entire pattern dictionary $P$:
\begin{equation} \label{e.problem}
	f(G,H) = \argmax_P \sum_i H(P^{(i)})
\end{equation}

The direct approach to solving for $f$ would require enumerating over all possible subgraphs of $G$, a computationally intractable task in most real-world scenarios since it would require storing the entirety of the graph stream, in addition to computing subgraph isomorphism checks over the entire graph. Therefore, we employ a heuristic algorithm to approximate such a solution.

\begin{algorithm}[t!]
\caption{GraphZip}
\label{alg:graphzip}
\begin{algorithmic}[1]
\STATE{Initialize $P$}
\WHILE{edges remain in stream}
	\STATE\label{line3}{Construct graph $B$ using \batchsize{} edges}
	\FOR{\textbf{each} graph $p$ in $P$}
		\STATE{$E$ $\gets$ subgraph isomorphisms of $p$ in $B$}
        \FOR{\textbf{each} graph $g$ in $E$}
        	\STATE{$g' \gets g.copy()$}
            \FOR{\textbf{each} $e$ in $E_g$}
                \IF{$e$ not in $p$}
                    \STATE{Extend $g'$ by new edge $e$}
                \ELSE
                	\STATE{Add internal edge $e$ to $g'$}
                \ENDIF                
            \ENDFOR
            \STATE{Mark each extended edge $e \in B$ as used}
            \IF{$g' \neq g$}
            	\STATE{Add $g'$ to $P$}
            \ENDIF
        \ENDFOR
	\ENDFOR
    \STATE{$R \gets$ remaining unused edges in $B$}
    \STATE{Add edges in $R$ as single-edge patterns to $P$}
\ENDWHILE
\RETURN $P$
\end{algorithmic}
\end{algorithm}

\subsection{The \system{\name} Algorithm}
\system{\name} is a highly-scalable method for discovering interesting patterns in a massive graph. Inspired by dictionary-based file compression, \system{\name} builds a dictionary of highly-compressing patterns by counting previously seen patterns in the graph stream and saving new patterns that extend from old ones. The resulting dictionary contains highly-compressing patterns from the given graph stream, which can be used directly or fed into a separate non-streaming algorithm (e.g., a maximal-clique finder). 
While \system{\name} is designed specifically with graph streams in mind, 
the algorithm can be easily applied to smaller static graphs without modification: 
if \system{\name} is given as input a single graph it will automatically partition it into batches of size \batchsize{} and process the graph as a stream. If the total number of edges in the graph (or number of edges remaining after $n$ iterations) is less than \batchsize{}, \system{\name} will process the graph as a single batch. This flexibility between input types allows us to compare \system{\name} directly with non-streaming methods.

\begin{figure}[t]
\includegraphics[width=\columnwidth]{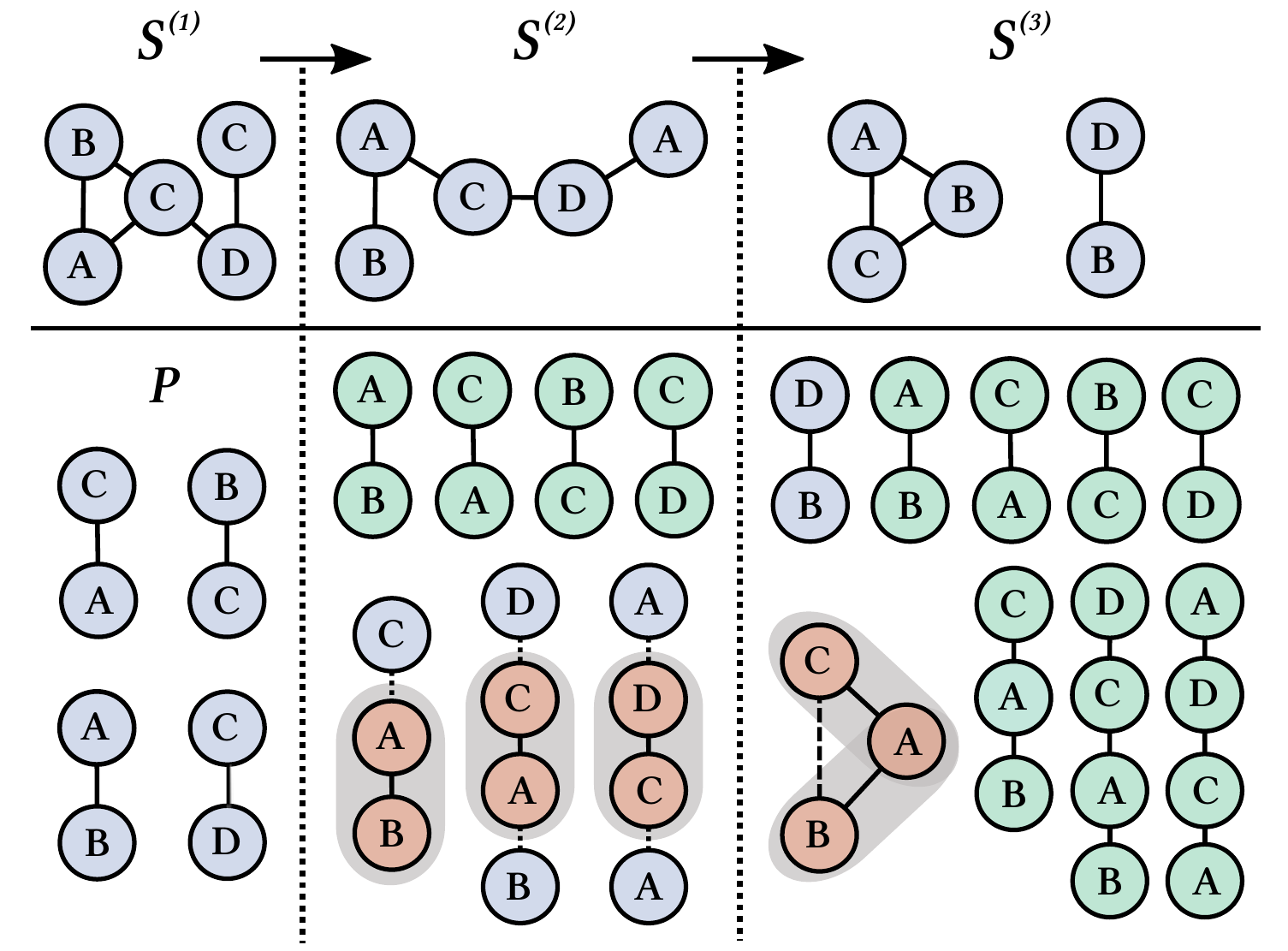}
\caption{A simplified illustration of the \system{\name} algorithm. In dictionary $P$, blue indicates a new pattern, orange indicates a matched pattern extending to a new pattern, and green indicates a non-repeated pattern. After processing $S^{(1)}$, $P$ contains only single edge patterns. $A$-$B$, $C$-$A$, and $D$-$C$ are embedded in $S^{(2)}$, so they are extended as new patterns in $P$. $C$-$A$-$B$ is embedded in $S^{(3)}$, and is extended with an internal edge as a new pattern, along with the remaining edge $D$-$B$.}
\label{fig:algorithm}
\end{figure}

The general procedure of \system{\name} is illustrated in figure 2. \system{\name} is initialized with an empty dictionary $P$ with max size \dictsize{} (provided by the user), which maps graphs to their frequency (count) and compression score.  \system{\name} collects arriving edges from the graph stream into batches of size \batchsize{} (also provided by the user), and runs the $compress$ procedure on each batch $B$: if a pattern $p$ from the dictionary is embedded in $B$, \system{\name} increments the frequency of the pattern in the dictionary and recomputes its compression score. Additionally, for each instance $i$ of pattern $p$ embedded in batch $B$, \system{\name} extends $p$ by one edge length, tagging each of the edges from $B$ used to extend $p$. The new edges used to extend $p$ are the edges incident on $i$ that exist in batch $B$ but not in pattern $p$. \system{\name} then adds the new extended pattern to $P$. After $P$ has been updated with all the extended patterns, the remaining untagged edges in $B$ are added as single-edge patterns to $P$. 
Our current reference implementation supports both undirected and directed edges, but not hyper-edges or self-loops. However, these limitations are implementation specific rather than inherent to the algorithm. Additionally, both representational variants can be converted to simple edges (a node and two edges).

If the dictionary exceeds size 2\dictsize{}, the dictionary is sorted according to the compression scores and trimmed to \dictsize{}. A pattern's compression score is computed as follows:
\begin{equation} \label{e.scorefunction}
	H(P^{(i)}) = (|E_{P^{(i)}}| - 1) \times (F_{P^{(i)}} - 1)
\end{equation}
This equates a pattern's compressibility to a product of its size and frequency. We use $(F_{P^{(i)}} - 1)$ so that a pattern with a frequency of $1$ has a compression score of $0$, since a pattern that only appears once affords no real compression  to the overall graph. The same offset is applied to the pattern size in $(|E_{P^{(i)}}| - 1)$ to reduce the weighting of single-edge patterns. 
Due to the fact that overlapping instances of a pattern in a batch are counted independently, it is possible that \system{\name} will overestimate the compression value of large structures with many homomorphisms. 

Note that the compression method used is intrinsically lossy, since \system{\name} does not retain information on how each of the instances are connected to the rest of the graph. The main focus of our work is knowledge discovery in graph streams, so lossy compression is an appropriate trade-off for decreased complexity and increased performance. More work is necessary to make \system{\name} lossless, for example in the case where it is necessary to fully reconstruct the original graph from the pattern dictionary.

See algorithm \ref{alg:graphzip} for pseudo-code, and the online repository for a reference implementation.

\subsection{Scalability}
Speed and memory usage are critical properties of graph mining algorithms designed to mine large real-world graphs. A deployed graph mining system should be able to keep up with the flow of data in the dynamic graph setting,
while summarizing a possibly infinite graph stream in memory. 
Memory usage in \system{\name} is directly bounded by the maximum dictionary size (\dictsize{}), 
and is indirectly bounded by the batch size (\batchsize{}), since the patterns within the dictionary cannot grow larger than the batch size (no subgraph isomorphisms of the pattern in the batch will exist). 
Both parameters \dictsize{} and \batchsize{} can be modified to maximize performance given certain hardware limitations.

The bulk of the computation in the \system{\name} algorithm happens while checking for embeddings of pattern $p$ in batch $B$ (\emph{find all subgraph isomorphisms of $p$ in $B$}). Note that because each entry in the pattern dictionary is unique, none of the subgraph isomorphism checks are contingent on each other, and thus the loop can be na\"{\i}vely parallelized across an arbitrary number of cores. This allows for large performance gains and means that an increase in dictionary size can be scaled linearly with an increase in cores. 
Even without parallelization of the subgraph isomorphism checks, \system{\name} is still faster than other state-of-the-art graph mining systems (as described in section \S 4). See section \S A for a formal runtime analysis.



\section{Experimental Evaluation}
There are two main questions we focus on when evaluating our 
algorithm: does it generate objectively and subjectively good results (i.e. correct and interesting results, respectively), and does it generate them in a reasonable amount of time? To answer these questions we test \system{\name} on an extensive suite of synthetic and real datasets ranging from a few thousand to several million nodes and edges (see table \ref{tab:datasets}). Using these datasets, we benchmark \system{\name} against three state-of-the-art, openly available graph mining systems: \system{SUBDUE}\footnote{\url{http://ailab.wsu.edu/subdue}}, \system{GraMi}\footnote{\url{https://github.com/ehab-abdelhamid/GraMi}}, and \system{StreamFSM}\footnote{\url{https://github.com/rayabhik83/StreamFSM}}.  
Because there is no directly comparable method to \system{\name} for mining maximally-compressing patterns in graph streams, we instead evaluate \system{\name} against a non-streaming method for mining compressing patterns (\system{SUBDUE}), a non-streaming method for frequent subgraph mining (\system{GraMi}), and a streaming method for frequent subgraph mining (\system{StreamFSM}). Highly-compressing patterns are often both large and frequent, so FSM methods serve as an appropriate comparison to \system{\name}. 

All experiments were run on a compute server configured with an AMD Opteron 6348 processor (2.8 GHz) and 128GB of RAM.

\begin{figure}[t!]
\includegraphics[width=\columnwidth]{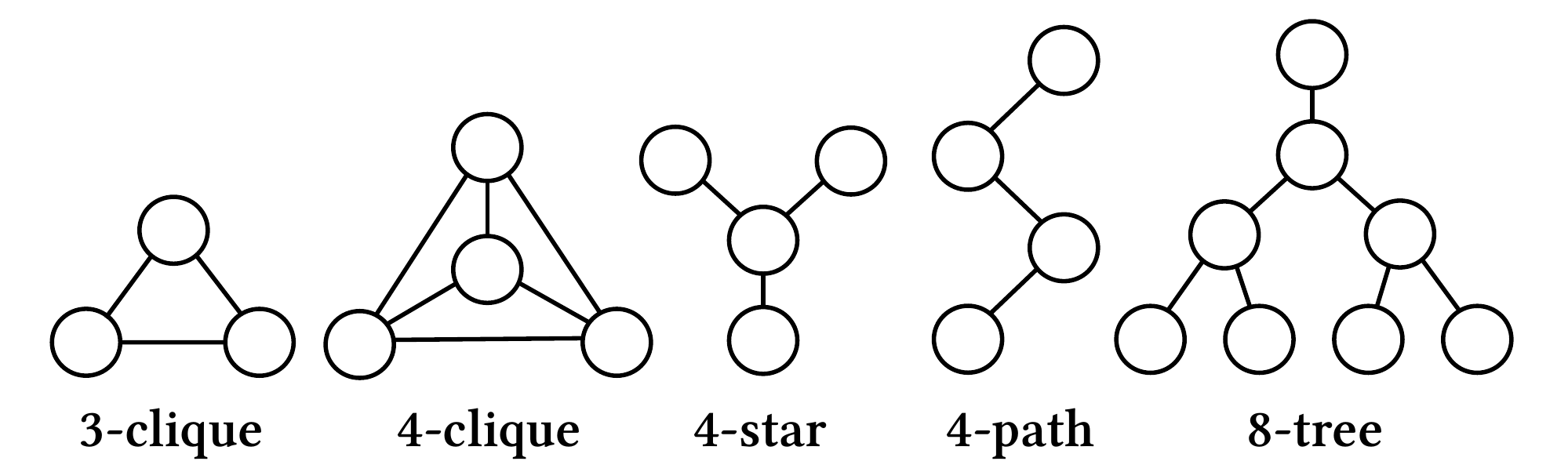}
\caption{We embed different types of graph substructures into our synthetic graphs, and then test to see if they are recovered. Corresponding datasets (from left to right): 3-CLIQ, 4-CLIQ, 4-STAR, 4-PATH, 8-TREE.}\label{fig:groups}
\end{figure}



\subsection{Synthetic graphs}
To test whether our algorithm outputs \emph{correct} substructures, we utilize a tool called \system{SUBGEN} \cite{subdue2-subgen} to embed ground truth patterns with desired frequencies into an artificially generated graph. This allows us to test whether or not a graph mining system correctly surfaces known patterns we expect to be returned in the result set. Since both \system{\name} and \system{SUBDUE} are designed to mine highly-compressing patterns from a graph, we embed large and frequent (i.e., highly-compressing) patterns in the graph, then record the number of ground truth patterns recovered. Given a set of embedded patterns $E$, and a set of patterns $R$ returned by our graph mining system, an embedded pattern $E^{(i)} \in E$ is considered \emph{matched} if for some returned pattern $R^{(i)} \in R$, $R^{(i)} \simeq E^{(i)}$.
Thus, we calculate the fraction $a$ of embedded patterns recovered using the scoring metric
\begin{equation} \label{e4.1}
	a = |\{E^{(i)} | E^{(i)} \in E \wedge R^{(i)} \in R \wedge E^{(i)} \simeq R^{(i)}\}|\ /\ |E|
\end{equation}
Which is equivalent to
\begin{equation} \label{e4.2}
    \textit{\small accuracy} = {\textit{\small matched patterns } } / { \textit{ \small total patterns}}
\end{equation}
We count a ground truth pattern as \emph{matched} if it is found in \system{\name}'s pattern dictionary after the final batch, or in \system{SUBDUE}'s case, if it is returned directly at the end of the program.

In addition to making the ground truth patterns highly-compressing, we also design the patterns to cover a wide class of fundamental graph patterns, including cliques, paths, stars and trees (see figure \ref{fig:groups}). This allows us to discern if a method has difficulty mining a certain type of structure (e.g., a poorly designed system may have trouble detecting cycles and therefore cliques). The naming scheme for each synthetic graph dataset is \emph{N-TYPE}, where $N$ is the number of vertices in the embedded pattern and \emph{TYPE} is a shorthand of the pattern type (e.g., 3-CLIQ is a graph with embedded 3-cliques). 
All synthetic graphs in table \ref{table:graphzip-subdue-synthetic} are generated with 1000 nodes, 5000 edges, and 20\%, 50\% or 80\% coverage (the percentage of the graph covered by instances of the pattern). 

\begin{table}[t]
  \caption{\system{\name} and \system{SUBDUE} runtime and accuracy on various synthetic graphs. For runtime, lower is better. For \system{SUBDUE}, the `{+}' for runtime indicates the program was terminated after 1000 seconds (no accuracy shown).}
  \label{table:graphzip-subdue-synthetic}
  \begin{tabular}{@{}lccc@{}c@{}cc@{}}
    \toprule
    \multirow{2}{*}{Dataset} & \multirow{2}{*}{Cov.} & \multicolumn{2}{c}{Runtime (sec.)} & \phantom{a} & \multicolumn{2}{c}{Accuracy (\%)} \\
    \cmidrule(lr{0em}){3-4} \cmidrule{6-7}
    & (\%) & {\small \system{\name}} & {\small \system{SUBDUE}} & \phantom{abc} & {\small \system{\name}} & {\small \system{SUBDUE}} \\
    \midrule
    \addlinespace[0.5em]
    \multirow{3}{*}{3-CLIQ} 
    & 20 & \textbf{52.25} & 66.68 & & \textbf{100.0} & 89.24 \\
    \addlinespace[-0.2em]
    & 50 & \textbf{3.779} & 22.22 & & \textbf{100.0} & 89.61 \\
	\addlinespace[-0.2em]
    & 80 & \textbf{3.665} & 11.99 & & \textbf{100.0} & 86.61 \\
	\addlinespace[0.3em]
    \multirow{3}{*}{4-PATH} 
    & 20 & \textbf{45.37} & 58.00 & & \textbf{100.0} & 100.0 \\  
    \addlinespace[-0.2em]
    & 50 & \textbf{3.052} & 18.57 & & \textbf{100.0} & 100.0 \\  
    \addlinespace[-0.2em]
	& 80 & \textbf{2.935} & 10.30 & & \textbf{100.0} & 100.0 \\  
    \addlinespace[0.3em]
    \multirow{3}{*}{4-STAR}
    & 20 & \textbf{50.70} & 1000{+} & & \textbf{100.0} & - \\
    \addlinespace[-0.2em]
    & 50 & \textbf{4.184} & 1000{+} & & \textbf{100.0} & - \\
    \addlinespace[-0.2em]
    & 80 & \textbf{4.483} & 1000{+} & & \textbf{100.0} & - \\
	\addlinespace[0.3em]
    \multirow{3}{*}{4-CLIQ}
    & 20 & \textbf{68.06} & 1000{+} & & \textbf{100.0} & - \\
    \addlinespace[-0.2em]
    & 50 & \textbf{29.19} & 1000{+} & & \textbf{100.0} & - \\
    \addlinespace[-0.2em]
    & 80 & \textbf{13.92} & 44.78 & & \textbf{100.0} & 89.51 \\
	\addlinespace[0.3em]
    \multirow{3}{*}{5-PATH}
    & 20 & \textbf{48.47} & 1000{+} & & \textbf{100.0} & - \\
    \addlinespace[-0.2em]
    & 50 & \textbf{4.461} & 21.30 & & \textbf{100.0} & 99.81 \\
    \addlinespace[-0.2em]
    & 80 & \textbf{4.267} & 24.16 & & \textbf{100.0} & 99.42 \\
	\addlinespace[0.3em]
    \multirow{3}{*}{8-TREE}
    & 20 & \textbf{62.68} & 1000{+} & & \textbf{100.0} & - \\
    \addlinespace[-0.2em]
    & 50 & \textbf{10.39} & 1000{+} & & \textbf{99.65} & - \\
    \addlinespace[-0.2em]
    & 80 & \textbf{11.07} & 1000{+} & & \textbf{100.0} & - \\
    \addlinespace[0.1em]
  	\bottomrule
\end{tabular}
\end{table}

\subsection{Comparison with \system{SUBDUE}}

Table \ref{table:graphzip-subdue-synthetic} shows the runtime and accuracy (eq. \ref{e4.1}) for \system{\name} and \system{SUBDUE} on the synthetic datasets. \system{\name} is clearly faster than \system{SUBDUE}, taking an order of magnitude less runtime in most experiments. Decreasing the coverage across all pattern types increased the runtime for both systems. 
\system{SUBDUE} is unable to process half of the datasets (including all 4-STAR and 8-TREE experiments) in less than 1000 seconds, while \system{\name} is able to process the same datasets in a fraction of the time with 99-100\% accuracy in all cases.

Among the datasets \system{SUBDUE} is able to process, the greatest difference in accuracy lies in the clique datasets (3-CLIQ and 4-CLIQ), where \system{SUBDUE} misses approximately 10\% of the embedded patterns. 
The \system{\name} algorithm contains an explicit edge case to extend internal edges in a pattern with no new vertices, which enables \system{\name} to capture cliques with high accuracy (see algorithm \ref{alg:graphzip} and figure \ref{fig:algorithm}). 

Our results indicate a stark difference in efficiency between \system{\name} and \system{SUBDUE}: \system{SUBDUE} is significantly slower than \system{\name} even on relatively small graphs with several thousand edges, and for graphs with certain classes of embedded patterns in them (stars and binary trees are particularly problematic). For this reason, when evaluating \system{\name} with larger real-world graphs we focus on benchmarking against more scalable methods.



\subsection{Real-world graphs}
We use several large, real-world graph datasets to test the scalability of the \system{\name} algorithm. See table \ref{tab:datasets} for further details. \\
\emph{NBER}\footnote{\url{http://www.nber.org/patents}}: NBER \cite{nber} is a graph of all U.S. patents granted (from Jan. 1963 to Dec. 1999) and the citations between them. The graph contains nearly 4 million nodes (patents) and over 16 million edges. Each node (citing patent) has edges to all the patents in its citation section. We added time-stamps to the citation graph prepared by \cite{snap-nber} and removed all withdrawn patents which had missing metadata ($<$ 0.04\% of all edges).\\
\emph{HetRec}\footnote{\url{http://grouplens.org/datasets/hetrec-2011}}: The HetRec 2011 MovieLens 2k \cite{hetrec} dataset links movies of the MovieLens 10M\footnote{\url{http://www.grouplens.org}} dataset with information from their IMDb\footnote{\url{http://www.imdb.com}} and Rotten Tomatoes\footnote{\url{http://www.rottentomatoes.com}} pages. We use a version of the dataset arranged by \cite{streamfsm}, in which nodes are labeled as `movie', `actor' or `director'. Edges connect movies to actors and directors: an edge from movie to director is labeled `directed-by', and an edge from movie to actor is labeled `acted-by'. The data spans 98 years, and is split into one graph stream (batch) file per year. \\
\emph{Higgs}\footnote{\url{https://snap.stanford.edu/data/higgs-twitter.html}}: The \emph{Higgs Twitter Dataset} \cite{higgs} is a collection of 
563,069 interactions (retweets, mentions, and replies) between 
304,691 users on Twitter before, during, and after the announcement of the discovery of Higgs boson particle on July 4th, 2012. The Tweets were scraped over the course of one week (168 hours) by filtering tweets for the tags `lhc', `cern', `boson' and `higgs'. 
Edges are labeled using the type of the interaction between the two users (`retweet', `mention', and `reply'), while all nodes share the same `user' label.

\begin{table}[t]
  \caption{Details of real-world datasets used. The average stream-rate (in edges per second) is calculated by dividing the total number of edges by the time span of the entire graph.
  }
  \label{tab:datasets}
  \begin{tabular}{@{}lrrrr@{}}
  	\toprule
    Dataset & Vertices & Edges & Labels & Stream-rate  \\
    \midrule
    NBER & 3,774,218 & 16,512,783 & 418 & $1.4 \times 10^{-2}$ \\
    Higgs & 304,691 & 563,069 & 4 & $9.3 \times 10^{-1}$  \\
    HetRec & 108,451 & 241,897 & 5 & $7.8 \times 10^{-5}$ \\
    \bottomrule
  \end{tabular}
\end{table}


\begin{figure*}[t]
\begin{minipage}[t]{0.48\linewidth}

\begin{figure}[H]
\centering

\begin{subfigure}[H]{0.47\linewidth}
    \includegraphics[width=\linewidth]{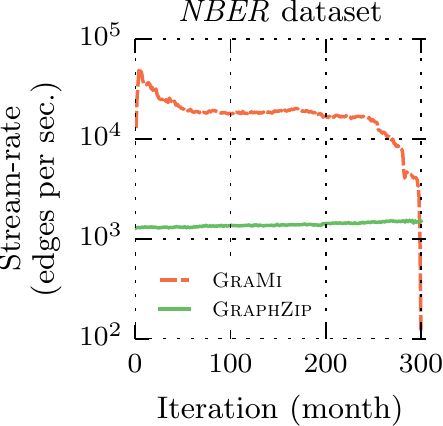}
    \caption{
    }
    \label{fig:graphzip_grami_sr_nber}
\end{subfigure}\qquad
\begin{subfigure}[H]{0.44\linewidth}
    \includegraphics[width=\linewidth]{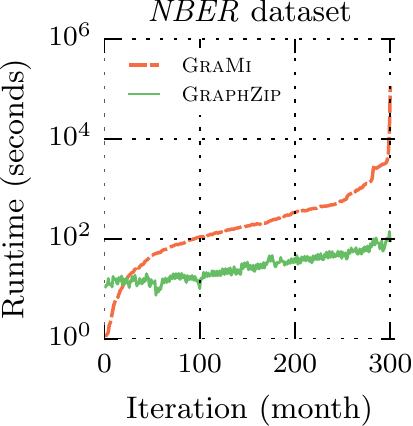}
    \caption{
    }
    \label{fig:graphzip_grami_rt_nber}
\end{subfigure}
\vspace{-3mm}
\caption[Two numerical solutions]{Stream-rate (a) and runtime (b) of \system{\name} and \system{GraMi} on the \textit{NBER} dataset. \system{GraMi}'s stream-rate decrease significantly near the end, while \system{\name}'s stream-rate remains relatively constant.}
\label{fig:graphzip_grami_nber}
\end{figure}

\end{minipage}%
    \hfill%
\begin{minipage}[t]{0.48\linewidth}

\begin{figure}[H]
\centering

\begin{subfigure}[H]{0.47\linewidth}
    \includegraphics[width=\linewidth]{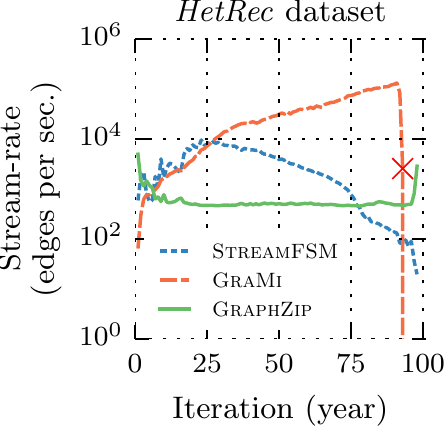}
    \caption{
    }
    \label{fig:graphzip_streamfsm_sr_hetrec_l}
\end{subfigure}\qquad
\begin{subfigure}[H]{0.44\linewidth}
    \includegraphics[width=\linewidth]{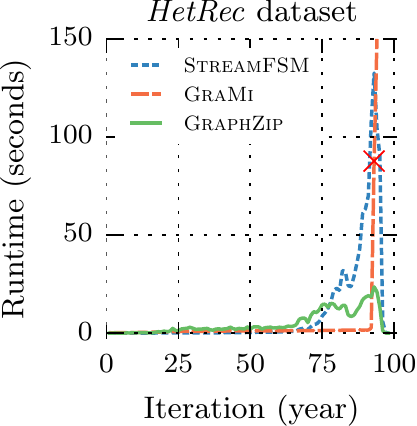}
    \caption{
    }
    \label{fig:graphzip_streamfsm_rt_hetrec_nl}
\end{subfigure}
\vspace{-3mm}
\caption[]{Stream-rate (a) and runtime (b) of \system{\name}, \system{GraMi}, and \system{StreamFSM} on the \textit{HetRec} dataset. Both \system{GraMi} and \system{StreamFSM} experience a massive spike in runtime near iteration 93.}
\label{fig:graphzip_streamfsm_hetrec}
\end{figure}

\end{minipage}%
\vspace{-1.0mm}
\end{figure*}

\begin{figure}
\vspace{2mm}
\centering
\begin{subfigure}[H]{0.45\linewidth}
    \includegraphics[width=\linewidth]{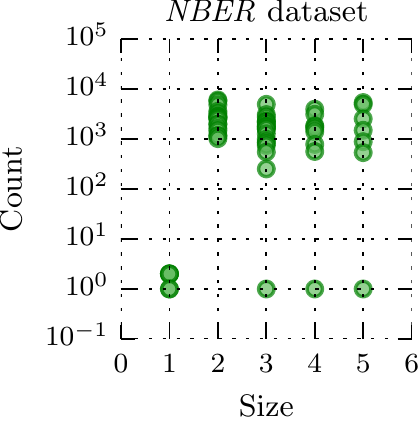}
    \caption{
    }
    \label{fig:graphzip_nber_scat}
\end{subfigure}\qquad
\begin{subfigure}[H]{0.45\linewidth}
    \includegraphics[width=\linewidth]{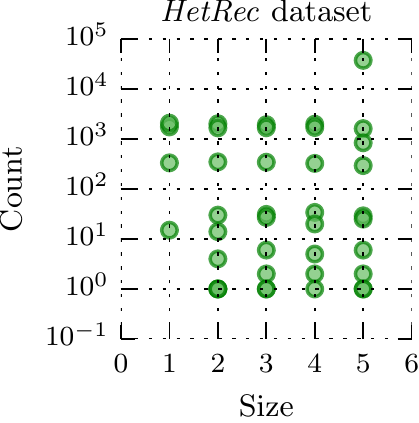}
    \caption{
    }
    \label{fig:graphzip_hetrec_scat}
\end{subfigure}
\vspace{-3mm}
\caption[]{Distribution of \system{\name}'s pattern dictionary after 300 iterations on \emph{NBER} (a) and after 98 iterations on \emph{HetRec} (b). The distribution for the larger \emph{NBER} dataset (a) is skewed towards patterns with high frequencies, while the distribution for patterns in \emph{HetRec} (b) is more varied.}
\label{fig:graphzip_scats}
\vspace{-1mm}
\end{figure}


\subsection{Comparison with \system{GraMi}}
Despite being designed for mining highly-compressing patterns like \system{\name}, \system{SUBDUE} has clear performance issues that restrict benchmarking it against \system{\name} on large real-world graphs. 
Therefore we also compare \system{\name} with \system{GraMi}, a state-of-the-art graph mining system for frequent subgraph mining on large static graphs. In contrast to \system{SUBDUE}, \system{GraMi} is efficient enough to process datasets with millions of edges, 
however \system{GraMi}'s relative performance still allows us to motivate the need for graph mining algorithms designed explicitly to handle streaming data.

\system{GraMi} takes as input a single graph file as opposed to a sequence of edges, so in order to simulate mining a dynamic graph with a non-streaming method we append the previous graph with the next set of edges at each iteration, initializing the graph with the first set of edges. Thus, each iteration represents a `snapshot' of the growing graph. Since both methods are paramaterized, we first tune \system{GraMi}'s minimum frequency threshold so that it returns a usable number of non-single edge patterns, then set \system{\name}'s parameters (batch and dictionary size) such that the pattern dictionary resembles the set of subgraphs returned by \system{GraMi}. On \emph{NBER} (figure \ref{fig:graphzip_grami_nber}), we fix \system{GraMi}'s minimum frequency threshold to 1000 which returned a set of subgraphs with a maximum, minimum, and average size of $6$, $1$, and $1.47$ respectively. Running \system{\name} with $\alpha = 5$ and $\theta = 50$ resulted in a pattern dictionary with a maximum, minimum, and average subgraph size of $5$, $1$, and $2.89$ (figure \ref{fig:graphzip_scats} shows the dictionary distributions for both \emph{NBER} and \emph{HetRec}). Since the overall runtime of each model depends significantly on the configuration of the parameters, 
the main purpose of our comparison is to examine trends in the runtime and stream-rate of each model using settings where they return comparable sets of subgraphs.

Our results show that \system{\name} is clearly more scalable than \system{GraMi} when mining large graphs in the streaming setting. While processing \emph{NBER}, \system{GraMi}'s runtime (figure \ref{fig:graphzip_grami_rt_nber}) grows exponentially, experiencing a large spike near iteration 300. Figure \ref{fig:graphzip_grami_sr_nber} (normalized by 
patents per month) demonstrates this clearly: \system{\name} maintains a constant stream-rate throughout, while \system{GraMi}'s stream-rate gradually slows until it sharply drops near the final updates. In fact, \system{\name}'s stream-rate  shows a slight increase over time; one explanation is that as the captured patterns in $P$ become more complex, less isomorphism checks occur per batch.

Results on the \emph{HetRec} dataset indicate similar trends, though to a more extreme degree. With \emph{HetRec}, we use a minimum frequency threshold of 9,000 for \system{GraMi} and keep the previous settings for \system{\name}: setting the threshold to 1,000 causes GraMi's stream-rate to slow to a relative crawl, and when using 10,000, \system{GraMi} is able to process the entire dataset but only returns two frequent subgraphs. While processing \emph{HetRec} with the threshold set to 9,000, \system{GraMi} maintains a high stream-rate which trends upwards over time until the 93rd iteration, where the system freezes and is unable to make any progress despite being left running for multiple days (as indicated by the red `X' on figures \ref{fig:graphzip_streamfsm_sr_hetrec_l} and \ref{fig:graphzip_streamfsm_rt_hetrec_nl}). 

\subsection{Comparison with \system{StreamFSM}}

Since there are no algorithms for mining highly-compressing subgraphs from graph streams in the existing literature, we benchmark \system{\name} against \system{StreamFSM}, a recently developed streaming algorithm for frequent subgraph mining. 
Subgraphs that compress well are often both frequent and large, so the tasks of mining highly-compressing and frequently-occurring subgraphs are closely related. 
The \system{StreamFSM} reference implementation available online was unable to find any frequently occurring subgraphs with any large datasets other than the provided \emph{HetRec} dataset (we hypothesize this is likely due to an implementation error), so we report results for \system{StreamFSM} on the HetRec dataset only.

A reasonable amount of time in the streaming setting equates to processing time less than or equal (at the very most) to the streaming-rate of the data; if the system cannot process the stream at the speed it is being generated, then the system is much less applicable in the real-life setting. Our results indicate that \system{\name} is significantly more scalable than \system{StreamFSM}: while \system{StreamFSM}'s stream-rate experiences an initial speedup, it quickly and consistently deteriorates after iteration 25, drastically increasing the runtime per iteration. The severe increase in runtime occurs around the same iteration that \system{GraMi} freezes (see figure \ref{fig:graphzip_streamfsm_rt_hetrec_nl}). In contrast, \system{\name} is seemingly unaffected by the same updates that cause massive slowdowns in \system{GraMi} and \system{StreamFSM}. \system{\name}'s stream-rate becomes relatively constant after an initial slowdown, and remains constant through to the end of the experiment (see figure \ref{fig:graphzip_streamfsm_sr_hetrec_l}). In the case of \system{\name} and \system{StreamFSM}, the stream-rate of both systems is much faster than the average stream-rate of the data ($7.8 \times 10^{-5}$ edges per second), despite \system{StreamFSM}'s relative volatility. However, a constant stream-rate is crucial for a deployed system processing a graph in real-time, since constraints on data processing time require predictable performance. 

\begin{figure}[t]
\centering
   \begin{subfigure}{\columnwidth}
\includegraphics[width=\columnwidth]{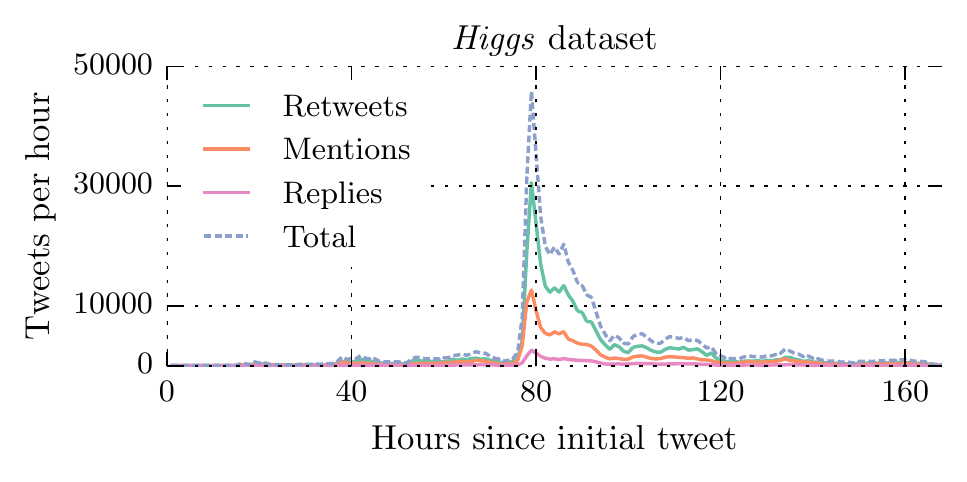}
\caption{Tracking Tweets per hour about the Higgs boson over time.}\label{fig:growth_higgs}
\end{subfigure}

\begin{subfigure}{\columnwidth}
\vspace{2mm}
\includegraphics[width=\columnwidth]{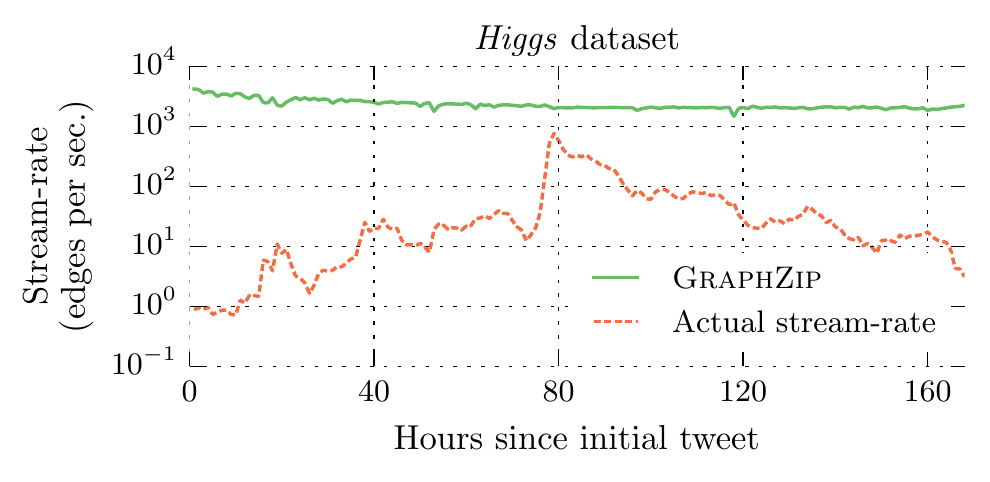}
\caption{Stream-rate of \system{\name} on the \emph{Higgs} dataset (higher is better).}\label{fig:sr_higgs}
\end{subfigure}
\vspace{-1mm}
\caption[]{(a) A large spike in activity occurs in the network after about 80 hours after the first Tweet. (b) After an initial slowdown, \system{\name} converges to a constant stream-rate.}
\end{figure}

\section{Twitter \& the Higgs boson particle}




One weakness of the datasets analyzed in the previous sections is the low granularity of their timestamps, e.g., \emph{HetRec} can only be split into real-time streaming units as small as year, and the synthetic datasets have no time information at all. 
Streaming intervals (and therefore time between results) as long as a year are unlikely in a real deployment setting, especially when disk space is taken into consideration (storing a year's worth of data before processing largely negates the benefit of streaming). For example, given a graph mining system configured to mine activity from a live network such as Twitter, it is likely that the user(s) would configure the interval to analyze patterns and trends over days, hours or even seconds. Additionally, reducing the time period between batches can reveal ebbs and flows in network activity that would be hidden by averaging out activity over a longer period.

The \emph{Higgs}'s dataset has time data in seconds for each interaction, so we are able to pre-process the dataset into graph files segmented by the hour.
One benefit to using the \emph{Higgs} dataset is to observe how large spikes in network traffic affect the stream-rate; the minimum, maximum, and average number of edges streamed per hour are 43, 45,861, and 3,352 respectively, with the peak number of tweets per hour coinciding with the official announcement of the discovery.

As we can see in figure \ref{fig:sr_higgs}, \system{\name}'s stream-rate is unaffected by the large spike in network traffic (using the same model parameters as the previous experiments). After an initial slowdown (similar to \emph{HetRec}), \system{\name}'s stream-rate converges on a constant stream-rate slightly faster than the maximum stream-rate the network reaches at the 80 hour mark, and much faster than the average stream-rate of the network ($9.3 \times 10^{-1}$ tweets per second). Our results indicate that if \system{\name} had been deployed to monitor the graph stream in real-time, it would have been able to process each set of updates before the next set of updates arrived.



\section{Conclusion and Future Work}
In this paper, we introduced \system{\name}, a graph mining algorithm that utilizes a dictionary-based compression approach to mine highly-compressing subgraphs from a graph stream. 
We showed that \system{\name} is able to successfully mine artificially-generated graphs for maximally-compressing patterns with comparable accuracy and much greater speed than a state-of-the-art approach. Additionally, we also demonstrated that \system{\name} is able to surface both complex and insightful patterns from large real-world graphs at speeds much faster than the actual stream-rate, 
with performance exceeding that of openly available state-of-the-art non-streaming and streaming methods. Future work will focus on implementing the potential optimizations to the algorithm discussed in this paper, including approximation algorithms for (subgraph) isomorphism computations and na\"{\i}ve parallelization.

\appendix
\section{Complexity Analysis}
In this section we examine the time complexity of algorithm \ref{alg:graphzip} 
in detail. We begin by analyzing the runtime per batch $B$ from a stream of graph $G$. Given a batch $B$, for each pattern $P^{(i)} \in P$ we retrieve the set $M$ of all embeddings (subgraph isomorphisms) of $P^{(i)}$ in $B$:
\begin{equation} \label{e.c1}
 \mathcal{O}(|P| \times \mathcal{O}(SI(P^{(i)},B)))
\end{equation}
Then, for each embedding $M^{(i)} \in M$, we (a) extend a copy of $M^{(i)}$ by one edge length in each direction and (b) add the new pattern to the dictionary:
\begin{equation} \label{e.c2}
\begin{aligned}
\mathcal{O}(&|P| \times (\mathcal{O}(SI(P^{(i)},B))\ + \\
& |M| \times (|V_{M^{(i)}}| \times k + |P| \times \mathcal{O}(I(M^{(i)'},P^{(i)})))))
\end{aligned}
\end{equation}
where $k$ is the number of edges incident of $v \in V_B$ shared with $u \in V_{M^{(i)}}$, and $M^{(i)'}$ is the extended copy of $M^{(i)}$. Finally, we add the remaining edges in $B$ as single-edge patterns ($e$) to $P$:
\begin{equation} \label{e.c3}
\begin{aligned}
\mathcal{O}(&|P| \times (\mathcal{O}(SI(P^{(i)},B))\ + \\
&|M| \times (|E_{M^{(i)}}| \times k + |P| \times \mathcal{O}(I(M^{(i)'},P^{(i)}))))\ +\\
&|R| \times \mathcal{O}(I(e,P^{(i)})))
\end{aligned}
\end{equation}
If we use the well-known VF2 algorithm to implement the subgraph and graph isomorphism functions, the time complexity for $SI(.)$ and $I(.)$ simplify to $\mathcal{O}(V^2)$ in the best case and $\mathcal{O}(V! \times V)$ in the worst case, where $V$ is the maximum number of vertices between the two graphs. Recall that our model is parameterized by $\theta$ and $\alpha$ ($|P|$ and $|E_B|$ respectively), which directly bounds $k < |V_B| < 2\alpha$, $|M| < 2^{\alpha}$ (maximum number of possible subgraphs in $B$), $E_{M^{(i)}} < \alpha$, and $|R| < \alpha$. Substituting in the worst case using VF2, we get:
\begin{equation} \label{e.c4}
\begin{aligned}
\mathcal{O}(&\theta \times (({(2\alpha)! \times 2\alpha)} \ + \\
&2^{\alpha} \times (\alpha \times 2\alpha + \theta \times {((2\alpha)! \times 2\alpha)}))\ +\\
&\alpha \times 2! \times 2)
\end{aligned}
\end{equation}
Which simplifies to:
\begin{equation} \label{e.c5}
\begin{aligned}
\mathcal{O}(&\theta \times \alpha! \times \alpha \ + \\
&\theta \times(2^{\alpha} \times \alpha^2 + 2^{\alpha} \times \theta \times \alpha! \times \alpha) +\\
&\theta \times \alpha)
\end{aligned}
\end{equation}
And lastly:
\begin{equation} \label{e.c6}
\mathcal{O}(\theta^2 \times 2^{\alpha} \times \alpha! \times \alpha)\\
= \mathcal{O}(\theta^2 \times \alpha!)
\end{equation}

Since $\theta$ and $\alpha$ are provided as parameters (constants at run-time), eq. \ref{e.c6} theoretically collapses to constant time ($\mathcal{O}(1)$), meaning that the \system{\name} algorithm scales linearly with the size of the overall graph. However, our complexity analysis illustrates an important trade-off in selecting the batch size and dictionary size, since too large a value for either parameter can exponentially increase the run-time per batch (increasing \batchsize{} is particularly costly).

Our theoretical complexity analysis reinforces our experimental findings: that subgraph isomorphism calculations (a known NP-complete problem) dwarf all other computations in the algorithm. A promising area of future work would be to incorporate existing approximation algorithms for the subgraph isomorphism problem to increase the efficiency of \system{\name}.


\begin{acks}
This material is based upon work supported by the \grantsponsor{}{National Science Foundation}{} under Grant No.:~\grantnum{}{1460917} and~\grantnum{}{1646640}.


\end{acks}

\bibliographystyle{abbrv}
\bibliography{long-refs} 

\begin{thebibliography}{10}

\bibitem{aggarwalvldb2010}
C.~C. Aggarwal, Y.~Li, P.~S. Yu, and R.~Jin.
\newblock On dense pattern mining in graph streams.
\newblock {\em {PVLDB}}, 3(1):975--984, 2010.

\bibitem{gmsurvey}
C.~C. Aggarwal and H.~Wang, editors.
\newblock {\em Managing and Mining Graph Data}, volume~40 of {\em Advances in
  Database Systems}.
\newblock Springer, 2010.

\bibitem{aggarwalicde2011}
C.~C. Aggarwal, Y.~Zhao, and P.~S. Yu.
\newblock Outlier detection in graph streams.
\newblock In {\em {ICDE}}, pages 399--409. {IEEE} Computer Society, 2011.

\bibitem{apriori}
R.~Agrawal and R.~Srikant.
\newblock Fast algorithms for mining association rules in large databases.
\newblock In {\em {VLDB}}, pages 487--499. Morgan Kaufmann, 1994.

\bibitem{giraph}
C.~Avery.
\newblock Giraph: Large-scale graph processing infrastructure on hadoop.
\newblock {\em Proceedings of the Hadoop Summit. Santa Clara}, 11, 2011.

\bibitem{germ}
M.~Berlingerio, F.~Bonchi, B.~Bringmann, and A.~Gionis.
\newblock Mining graph evolution rules.
\newblock In {\em {ECML/PKDD}}, pages 115--130. Springer, 2009.

\bibitem{braun2014}
P.~Braun, J.~J. Cameron, A.~Cuzzocrea, F.~Jiang, and C.~K. Leung.
\newblock Effectively and efficiently mining frequent patterns from dense graph
  streams on disk.
\newblock {\em Procedia Computer Science}, 35:338--347, 2014.

\bibitem{bron-kerbosch-cliques}
C.~Bron and J.~Kerbosch.
\newblock Finding all cliques of an undirected graph (algorithm 457).
\newblock {\em Commun. {ACM}}, 16(9):575--576, 1973.

\bibitem{hetrec}
I.~Cantador, P.~Brusilovsky, and T.~Kuflik.
\newblock Second workshop on information heterogeneity and fusion in
  recommender systems (hetrec2011).
\newblock In {\em RecSys}, pages 387--388. {ACM}, 2011.

\bibitem{ChingFB}
A.~Ching, S.~Edunov, M.~Kabiljo, D.~Logothetis, and S.~Muthukrishnan.
\newblock One trillion edges: Graph processing at facebook-scale.
\newblock {\em {PVLDB}}, 8(12):1804--1815, 2015.

\bibitem{subdue2-subgen}
D.~J. Cook and L.~B. Holder.
\newblock Substructure discovery using minimum description length and
  background knowledge.
\newblock {\em J. Artif. Intell. Res. {(JAIR)}}, 1:231--255, 1994.

\bibitem{subdue4}
D.~J. Cook and L.~B. Holder.
\newblock Graph-based data mining.
\newblock {\em {IEEE} Intelligent Systems}, 15(2):32--41, 2000.

\bibitem{subdue3}
D.~J. Cook, L.~B. Holder, and S.~Djoko.
\newblock Knowledge discovery from structural data.
\newblock {\em J. Intell. Inf. Syst.}, 5(3):229--248, 1995.

\bibitem{countminsketch}
G.~Cormode and S.~Muthukrishnan.
\newblock An improved data stream summary: the count-min sketch and its
  applications.
\newblock {\em J. Algorithms}, 55(1):58--75, 2005.

\bibitem{higgs}
M.~De~Domenico, A.~Lima, P.~Mougel, and M.~Musolesi.
\newblock The anatomy of a scientific rumor.
\newblock {\em (Nature Open Access) Scientific Reports}, 3, 2013.

\bibitem{grami}
M.~Elseidy, E.~Abdelhamid, S.~Skiadopoulos, and P.~Kalnis.
\newblock {GRAMI:} frequent subgraph and pattern mining in a single large
  graph.
\newblock {\em {PVLDB}}, 7(7):517--528, 2014.

\bibitem{subiso-npcomplete}
M.~R. Garey and D.~S. Johnson.
\newblock {\em Computers and intractability}, volume~29.
\newblock WH Freeman New York, 2002.

\bibitem{seus}
S.~Ghazizadeh and S.~S. Chawathe.
\newblock Seus: Structure extraction using summaries.
\newblock In {\em Discovery Science}, volume 2534 of {\em Lecture Notes in
  Computer Science}, pages 71--85. Springer, 2002.

\bibitem{nber}
B.~H. Hall, A.~B. Jaffe, and M.~Trajtenberg.
\newblock The {NBER} patent citation data file: Lessons, insights and
  methodological tools.
\newblock Technical report, National Bureau of Economic Research, 2001.

\bibitem{subdue}
L.~B. Holder, D.~J. Cook, and S.~Djoko.
\newblock Substucture discovery in the {SUBDUE} system.
\newblock In {\em {KDD} Workshop}, pages 169--180. {AAAI} Press, 1994.

\bibitem{fsg}
M.~Kuramochi and G.~Karypis.
\newblock Frequent subgraph discovery.
\newblock In {\em {ICDM}}, pages 313--320. {IEEE} Computer Society, 2001.

\bibitem{grew}
M.~Kuramochi and G.~Karypis.
\newblock {GREW-A} scalable frequent subgraph discovery algorithm.
\newblock In {\em {ICDM}}, pages 439--442. {IEEE} Computer Society, 2004.

\bibitem{sigram}
M.~Kuramochi and G.~Karypis.
\newblock Finding frequent patterns in a large sparse graph.
\newblock {\em Data Min. Knowl. Discov.}, 11(3):243--271, 2005.

\bibitem{snap-nber}
J.~Leskovec, J.~M. Kleinberg, and C.~Faloutsos.
\newblock Graphs over time: densification laws, shrinking diameters and
  possible explanations.
\newblock In {\em {KDD}}, pages 177--187. {ACM}, 2005.

\bibitem{streamsurvey}
A.~McGregor.
\newblock Graph stream algorithms: a survey.
\newblock {\em {SIGMOD} Record}, 43(1):9--20, 2014.

\bibitem{przulj-motifs}
N.~Przulj.
\newblock Biological network comparison using graphlet degree distribution.
\newblock {\em Bioinformatics}, 26(6):853--854, 2010.

\bibitem{graphzip2}
Y.~Qian and K.~Zhang.
\newblock Graphzip: a fast and automatic compression method for spatial data
  clustering.
\newblock In {\em {SAC}}, pages 571--575. {ACM}, 2004.

\bibitem{graphsig}
S.~Ranu and A.~K. Singh.
\newblock Graphsig: {A} scalable approach to mining significant subgraphs in
  large graph databases.
\newblock In {\em {ICDE}}, pages 844--855. {IEEE} Computer Society, 2009.

\bibitem{streamfsm}
A.~Ray, L.~B. Holder, and S.~Choudhury.
\newblock Frequent subgraph discovery in large attributed streaming graphs.
\newblock In {\em BigMine}, volume~36 of {\em {JMLR} Workshop and Conference
  Proceedings}, pages 166--181. JMLR.org, 2014.

\bibitem{mdl}
J.~Rissanen.
\newblock {\em Sochastic Complexity in Statistical Inquiry}.
\newblock Singapore: World Scientific Publishing, 1989.

\bibitem{graphscope}
J.~Sun, C.~Faloutsos, S.~Papadimitriou, and P.~S. Yu.
\newblock Graphscope: parameter-free mining of large time-evolving graphs.
\newblock In {\em {KDD}}, pages 687--696. {ACM}, 2007.

\bibitem{tcm}
N.~Tang, Q.~Chen, and P.~Mitra.
\newblock Graph stream summarization: From big bang to big crunch.
\newblock In {\em {SIGMOD} Conference}, pages 1481--1496. {ACM}, 2016.

\bibitem{arabesque}
C.~H.~C. Teixeira, A.~J. Fonseca, M.~Serafini, G.~Siganos, M.~J. Zaki, and
  A.~Aboulnaga.
\newblock Arabesque: a system for distributed graph mining.
\newblock In {\em {SOSP}}, pages 425--440. {ACM}, 2015.

\bibitem{margin}
L.~T. Thomas, S.~R. Valluri, and K.~Karlapalem.
\newblock {MARGIN:} maximal frequent subgraph mining.
\newblock volume~4, pages 10:1--10:42, 2010.

\bibitem{doulion}
C.~E. Tsourakakis, U.~Kang, G.~L. Miller, and C.~Faloutsos.
\newblock {DOULION:} counting triangles in massive graphs with a coin.
\newblock In {\em {KDD}}, pages 837--846. {ACM}, 2009.

\bibitem{wackersreuther}
B.~Wackersreuther, P.~Wackersreuther, A.~Oswald, C.~B{\"o}hm, and K.~M.
  Borgwardt.
\newblock Frequent subgraph discovery in dynamic networks.
\newblock In {\em Proceedings of the Eighth Workshop on Mining and Learning
  with Graphs}, pages 155--162. ACM, 2010.

\bibitem{leap}
X.~Yan, H.~Cheng, J.~Han, and P.~S. Yu.
\newblock Mining significant graph patterns by leap search.
\newblock In {\em {SIGMOD} Conference}, pages 433--444. {ACM}, 2008.

\bibitem{gspan}
X.~Yan and J.~Han.
\newblock gspan: Graph-based substructure pattern mining.
\newblock In {\em {ICDM}}, pages 721--724. {IEEE} Computer Society, 2002.

\bibitem{closegraph}
X.~Yan and J.~Han.
\newblock Closegraph: mining closed frequent graph patterns.
\newblock In {\em {KDD}}, pages 286--295. {ACM}, 2003.

\bibitem{aggarwalicdm2013}
W.~Yu, C.~C. Aggarwal, S.~Ma, and H.~Wang.
\newblock On anomalous hotspot discovery in graph streams.
\newblock In {\em {ICDM}}, pages 1271--1276. {IEEE} Computer Society, 2013.

\bibitem{zhaoicde2016}
P.~Zhao, C.~C. Aggarwal, and G.~He.
\newblock Link prediction in graph streams.
\newblock In {\em {ICDE}}, pages 553--564. {IEEE} Computer Society, 2016.

\bibitem{gsketch}
P.~Zhao, C.~C. Aggarwal, and M.~Wang.
\newblock gsketch: On query estimation in graph streams.
\newblock {\em {PVLDB}}, 5(3):193--204, 2011.

\bibitem{lz78}
J.~Ziv and A.~Lempel.
\newblock Compression of individual sequences via variable-rate coding.
\newblock {\em {IEEE} Trans. Information Theory}, 24(5):530--536, 1978.

\end{thebibliography}

\end{document}